\documentclass[12pt,twoside,natbib,usegraphicx]{article}
\usepackage{xrb2007}
\usepackage{graphics}

\begin{document}

\session{Faint XRBs and Galactic LMXBs}

\shortauthor{Farrell, Barret, and Skinner} \shorttitle{
\emph{Swift} Super-Orbital X-ray Binaries}

\title{A \emph{Swift} BAT Look at Super-Orbital X-ray Binaries}
\vspace{-0.5cm}
\author{Sean A. Farrell, Didier Barret}
\affil{Centre
d'Etude Spatiale des Rayonnements, CNRS/UPS, 9 Avenue du Colonel
Roche, 31028 Toulouse Cedex 4, France}
\author{and Gerald Skinner\altaffilmark{1}}
\affil{CRESST and Astroparticle Physics Laboratory, Code 661,
NASA/Goddard Space Flight Center, Greenbelt, MD 20771, USA}
\altaffiltext{1}{Department of Astronomy, University of Maryland, College
Park, MD 20742, USA}

\begin{abstract}
We present the results of a study with the \emph{Swift} Burst
Alert Telescope in the 14 -- 195 keV range of the long-term
variability of 5 low mass X-ray binaries with reported or
suspected super-orbital periods --- 4U 1636-536, 4U 1820-303, 4U
1916-053, Cyg X-2 and Sco X-1. No significant periodic modulation
was detected around the previously reported values in the 4U
1916-053, Cyg X-2 or Sco X-1 light curves. The $\sim$170 d period
of 4U 1820-303 was detected up to 24 keV, consistent with the
proposed triple system model. The $\sim$46 d period in 4U 1636-536
was detected up to 100 keV, clearly inconsistent with variable
photoelectric absorption via a warped precessing disc. We
speculate that the appearance of this modulation after 4U 1636-536
entered the low/hard state indicates that this variability could
be linked to jet precession such as observed in SS 433.
\end{abstract}
\vspace{-0.5cm}
\section{Introduction}
Long-term ``super-orbital'' periodic variability has been observed
in soft X-rays from over 30 X-ray binaries with the All Sky
Monitor (ASM) on the \emph{RXTE} satellite
\citep[e.g.][]{Farrell_so07}. Also known as `long' or `third'
periods, these modulations are defined simply as any periodic
variability greater than the orbital period. The precession of a
warped accretion disc modulating the X-ray flux from the compact
object is currently the favoured model
\citep[e.g][]{Farrell_wij99}. \citet{Farrell_og01} showed that
while radiation driven warping gives a coherent picture of
super-orbital periods in some systems, this model cannot explain
the observed phenomena in all cases.

The limited spectral coverage of the \emph{RXTE} ASM has so far
precluded long-term studies of super-orbital variability at high
energies. The launch of \emph{Swift} in 2004 saw the arrival of a
wide field monitoring instrument capable of complementing the
\emph{RXTE} ASM in hard X-rays -- the Burst Alert Telescope
\citep[BAT;][]{Farrell_bar05}. In this paper we present the
results of a search for the previously reported super-orbital
modulation of 5 low mass X-ray binaries in the BAT 14 -- 195 keV
light curves.

\section{Data Analysis}

The BAT 14 -- 24 keV (A), 24 -- 50 keV (B), 50 -- 100 keV (C) and
100 -- 195 keV (D) light curves were used for these analyses,
after correction for the off-axis position and earth occultations.
For comparison, the \emph{RXTE} ASM 1.5 -- 12 keV 1 d average
light curves for the same time span were analysed to cover the low
energy range of the spectrum.

The modified weighting scheme developed by \citet*{Farrell_co07}
was used. In this method, individual data points are given weights
based on both the non-uniform measurement uncertainties and the
intrinsic source variability. Power spectra were generated for
each of the weighted light curves using the fast periodogram
\textsc{fasper} subroutine of the Lomb-Scargle periodogram
\citep[see][and references therein]{Farrell_pr89}, with the 99$\%$
white noise significance levels estimated using Monte Carlo
simulations \citep*[e.g.][]{Farrell_ko98}.

\section{Results and Discussion}

\begin{figure*}
\begin{center}
\scalebox{0.8}{\includegraphics{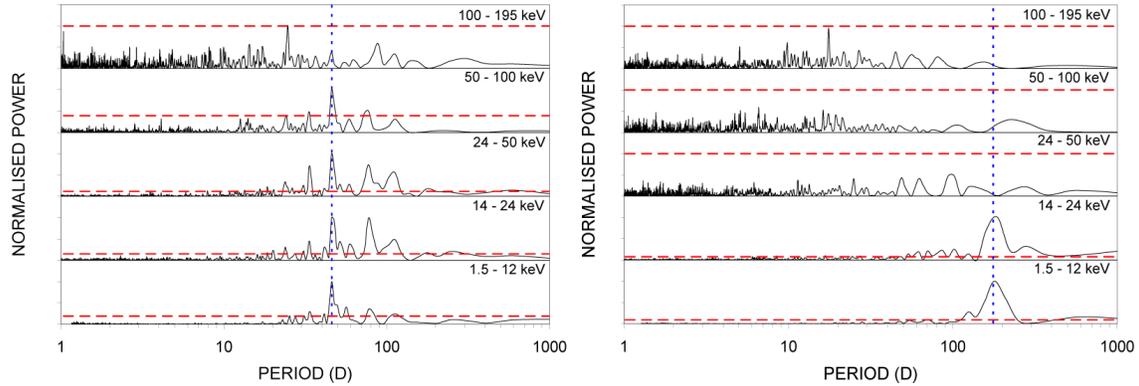}}
\caption{Normalised power spectra of the ASM and BAT light curves
of 4U 1636-536 (left panel) and 4U 1820-303 (right panel). The red
horizontal dashed lines indicate the 99$\%$ white noise
significance levels. The blue vertical dotted line indicates the
location of the reported super-orbital periods at $\sim$46 d (4U
1636-536) and $\sim$170 d (4U 1820-303) respectively.}\label{fig1}
\end{center}
\end{figure*}

No evidence for the reported super-orbital periods in 4U 1916-053
\citep[199 d;][]{Farrell_gr92}, Cyg X-2 \citep[77
d;][]{Farrell_sm92} or Sco X-1 \citep[37 d/62
d;][]{Farrell_pe96,Farrell_og01} was found in the ASM or any of
the BAT light curves. The power spectra were instead dominated by
significant low-frequency noise, with no peaks appearing to
indicate the presence of real long-term periodic modulation. The
super-orbital periods in 4U 1916-053 and Cyg X-2 were originally
detected using \emph{Vela 5B} data, while the detection of
super-orbital modulation in Sco X-1 was made using only the first
9 months of ASM data. It is thus apparent that either the reported
detections were spurious or the mechanisms behind the
super-orbital variability in each of these three systems has since
ceased. In comparison, the $\sim$46 d and $\sim$170 d variability
previously reported for 4U 1636-536 \citep{Farrell_sh05} and 4U
1820-303 \citep{Farrell_pr84b} respectively were detected in both
the ASM and BAT light curves (Figure \ref{fig1}).

The $\sim$170 d period in 4U 1820-303 has been attributed to
variable accretion linked to the presence of a third body in this
system, whereby the eccentricity of the inner orbit is modulated
by tidal effects produced by the outer body \citep{Farrell_cho01}.
Transitions between the high/soft and low/hard states over this
timescale are linked to variability in the accretion rate
\citep{Farrell_bl00b} and the movement in and out of the inner
region of the accretion disc \citep[see for
example][]{Farrell_mal07}, thus modulating the thermal
Comptonisation region of the spectrum. This thermal Comptonisation
component has been seen to dominate the spectrum up to $\sim$50
keV, above which a high energy non-thermal tail is dominant
\citep{Farrell_ta07}. The non-detection of the modulation above 24
keV is therefore due to a lack of sensitivity of the BAT (as the
modulation should be present in the 24 -- 50 keV band), precluding
us from testing the idea that the modulation also affects the
non-thermal tail and thus extends to higher energies.

The detection of the $\sim$46 d period in 4U 1636-536 up to 100
keV is clearly inconsistent with variable photoelectric absorption
by a warped precessing disc, as absorption should effect only the
low energy (ASM) data. \citet{Farrell_fi06} linked the recent
reduction in X-ray flux and the appearance of a high energy tail
detected by \emph{INTEGRAL} in the spectrum of 4U 1636-536 to the
onset of jet formation. The detection of the modulation at
energies where the high energy tail is dominant (i.e. $>$ 30 keV)
indicates that both the thermal Comptonisation and high energy
tail components are varying over this timescale. Coincidentally,
the $\sim$46 d modulation first appears in the ASM data around the
same time as this transition to the low/hard state. The appearance
of a high energy power law tail in the spectrum in conjunction
with the first detection of the $\sim$46 d period and the
reduction in flux is intriguing, implying a link between the three
phenomena. If the assertion made by \citet{Farrell_fi06} is
correct --- that the high energy tail is produced in compact jets
--- then the $\sim$46 d modulation is likely also associated in
some way with the jets. The modulation could thus be tied to
either transient jet formation or the precession of the jets
(linked in turn to the precession of a warped disc) in a similar
fashion as seen in SS 433 \citep{Farrell_ma84}. Further
observations at radio wavelengths are being sought in order to
confirm this hypothesis.\\

\acknowledgments
\footnotesize{This work made use of quick-look ASM results
provided by the ASM/\emph{RXTE} team. The authors thank members of
the \emph{Swift} BAT team for helpful discussions.}

\end{document}